\documentclass[prl,superscriptaddress,onecolumn]{revtex4-1}
\usepackage{times,amsmath}
\usepackage{epsfig}
\usepackage{color}
\usepackage{longtable}
\usepackage{graphicx}
\usepackage{dcolumn}
\usepackage{bm}
\usepackage{tabularx}
\usepackage{hyperref}
\usepackage{multirow}
\hypersetup{citecolor=black, filecolor= black, linkcolor= black, urlcolor= black}

\begin{document}
\preprint{version 0}

\title{Formation of Stoichiometric CsF$_n$ Compounds}
\author{Qiang Zhu}
\email{qiang.zhu@stonybrook.edu}
\affiliation{Department of Geosciences, Stony Brook University, Center for Materials by Design, Institute for Advanced Computational Science, Stony Brook University, NY 11794, USA}
\author{Artem R. Oganov}
\affiliation{Department of Geosciences, Stony Brook University, Center for Materials by Design, Institute for Advanced Computational Science, Stony Brook University, NY 11794, USA}
\affiliation{Department of Problems of Physics and Energetics, Moscow Institute of Physics and Technology, 9 Institutskiy lane, Dolgoprudny city, Moscow Region, 141700, Russia}
\affiliation{School of Materials Science and Engineering, Northwestern Polytechnical University, Xi'an,710072, China}
\author{Qingfeng Zeng}
\affiliation{Science and Technology on Thermostructural Composite Materials Laboratory, Northwestern Polytechnical University, Xi'an, 710072, China}
\date{\today}

\begin{abstract}
Alkali halides $MX$, have been viewed as typical ionic compounds, characterized by 1:1 ratio necessary for charge balance between M$^+$ and X$^-$. 
It was proposed that group I elements like Cs can be oxidized further under high pressure. 
Here we perform a comprehensive study for the CsF-F system at pressures up to 100 GPa, and find extremely versatile chemistry. 
A series of CsF$_n$ ($n$ $\geq$ 1) compounds are predicted to be stable already at ambient pressure. 
Under pressure, 5$p$ electrons in Cs atoms become active, with growing tendency to form Cs$^{3+}$ and Cs$^{5+}$ valence states at fluorine-rich conditions. 
Although Cs$^{2+}$ and Cs$^{4+}$ are not energetically favoured, the interplay between two mechanisms (polyfluoride anions and polyvalent Cs cations) allows CsF$_2$ and CsF$_4$ compounds to be stable under pressure. 
The estimated defluorination temperatures of CsF$_n$ (n=2,3,5) compounds at atmospheric pressure (218 $^\circ$C, 150 $^\circ$C, -15 $^\circ$C, respectively), are attractive for fluorine storage applications.
\end{abstract}

\maketitle

\section{Introduction}
In general, for a given ionic compound A$_m$B$_n$, the stoichiometry reflects the ratio of valences (i.e., the charges for each anion and cation). 
Yet, some ionic compounds do not strictly obey this rule. 
For instance, MgO$_2$ can be prepared at very high oxygen fugacities, in which anions form the peroxide group [O$_2$]$^{2-}$ \cite{Zhu-MgO-PCCP}. 
The variation of stoichiometry comes from the formation of polyatomic anions (such as those in peroxides, superoxides \cite{Peroxides-2007}, polyiodides \cite{Polyiodide-2003}, .etc), without changing cation valences. 
It was found that peroxides (usually unstable or metastable), become thermodynamically stable under higher pressures \cite{Zhu-MgO-PCCP}.

It appears that increasing pressure promotes the formation of increasing oxidation states. 
Our recent work has discovered that Xe will form stable compounds with O under high pressure, in which Xe exhibits high oxidation states of +2, +4, and +6 \cite{Zhu-NChem-2013}. 
In this case, the new stoichiometry is no longer from the anion-anion bonds (O-O), but from the increased valence of Xe.
Elements around Xe in the Periodic Table are expected to undergo similar transitions. 
In particular, Cs, in the electronic configuration [Xe]6$s$$^1$, is a natural choice to study this possibility. 
Indeed, Miao \cite{Miao-NChem-2013} has recently reported that Cs, under pressure, can adopt oxidation states higher than +1 to form a series of stable CsF$_n$ compounds. 
According to Miao's calculation, CsF$_2$ becomes stable at 5 GPa, CsF$_3$ at 15 GPa, and CsF$_5$ at 50 GPa. 
Considering [CsF$_2$]$^-$ and [CsF$_5$] are isoelectronic to the well-known molecular XeF$_2$ \cite{Levy-JACS-1963} and [XeF$_5$]$^-$ \cite{XeF5-1991}, this picture makes sense. 
However, Miao suggested CsF$_2$ and CsF$_4$ might adopt structures similar to those of XeF$_2$ and XeF$_4$, with the valence state of Cs being +2 and +4. 
In particular, an $I$4/$mmm$ CsF$_2$ was proposed to be thermodynamically stable at 10 GPa, which appears to break the isoelectronic analogy and involve Cs$^{2+}$ ions isoelectronic to unknown and unstable Xe$^+$ ion. 
To resolve this, we performed a comprehensive investigation of CsF$_n$ system under pressures up to 100 GPa. 
Our calculation uncovers quite a different scenario from Miao's report. 
We further explain that interplay between two mechanisms (polyfluoride anions and evolution of Cs valence state) creates an unexpected variety of stable CsF$_n$ compounds under moderate pressure.

\section{Results and Discussion}
\begin{figure}
\epsfig{file=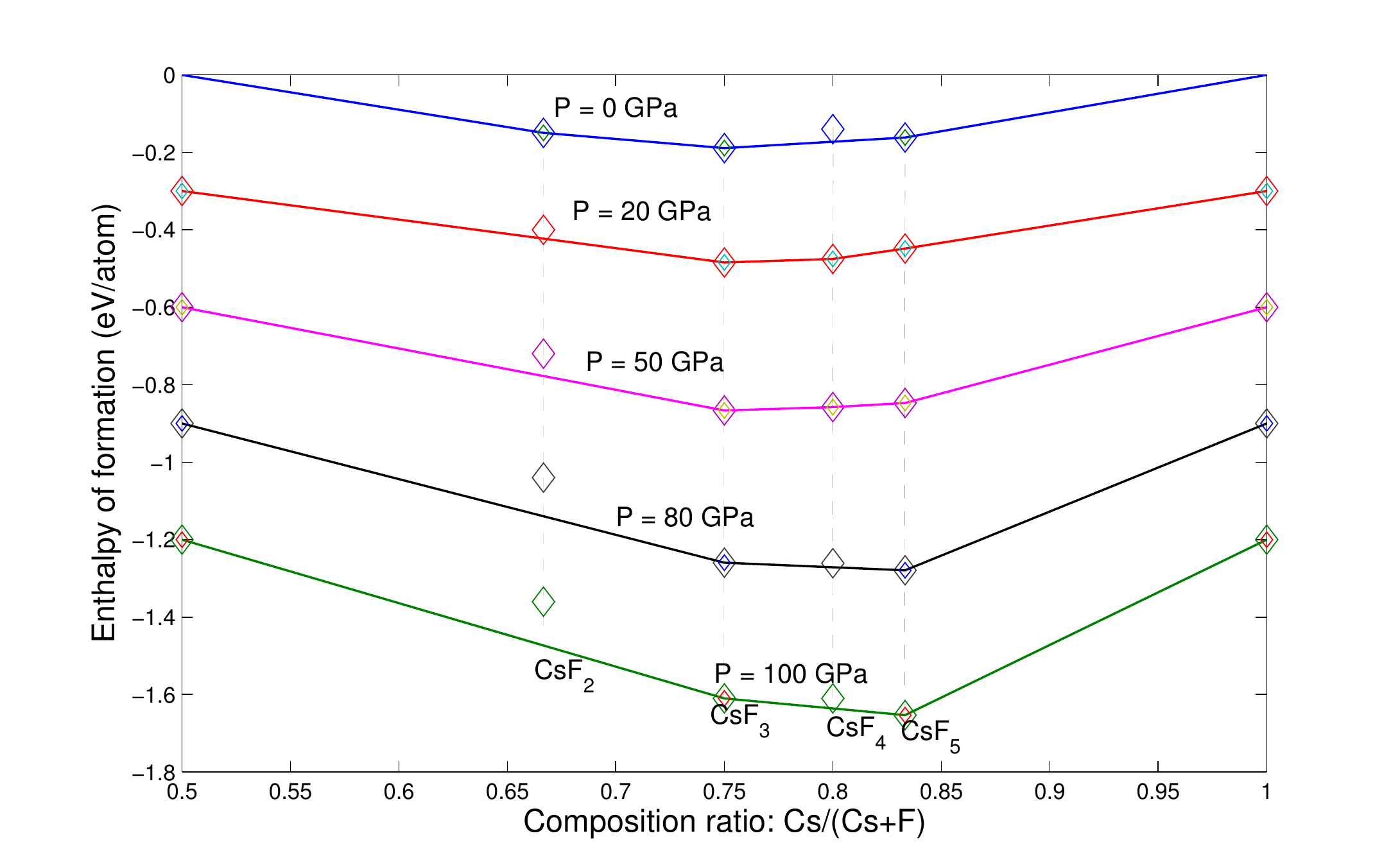, width=0.5\textwidth}
\caption{\label{hull} Convex hull diagrams of CsF$_n$ at different pressures.}
\end{figure}

We have performed variable-composition structure searches using the USPEX code \cite{Oganov-JCP-2006, Oganov-ACR-2011, Lyakhov-CPC-2013, Zhu-Acta-2012, Zhu-PRB-2013} with up to 24 atoms in the unit cell at pressures at 0 and 100 GPa for the Cs-F system, in which we found only CsF$_n$ ($n\geq$1) to be stable, with CsF being stable at all pressures. 
Thus we focused our search on the CsF-F system in the range of 0, 30, 50, 75, 100 GPa.
These searches yielded the correct crystal structures for CsF and F, and a series of polyfluoride compounds as stable states, including CsF$_2$, CsF$_3$, CsF$_5$ which are thermodynamically stable already at ambient pressure. 
As pressure increases, CsF$_2$ becomes unstable above 19 GPa, while CsF$_4$ appears on the convex hull between 17-80 GPa. 
CsF$_3$ and CsF$_5$ are stable in the entire pressure range up to 100 GPa. 
However, all of the stable compounds undergo a series of phase transitions, with dramatic changes of the electronic structure. 

\begin{figure}
\epsfig{file=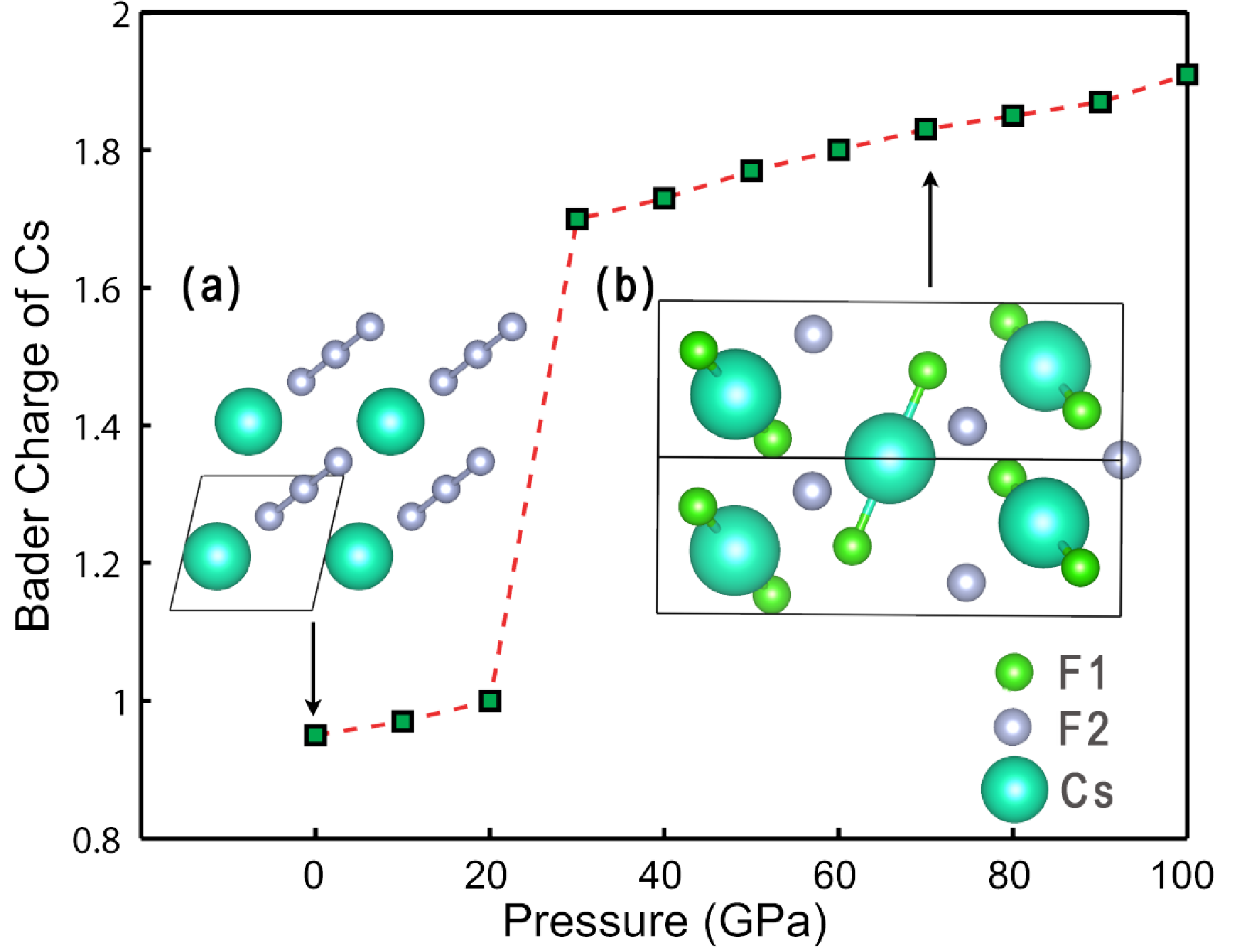, width=0.50\textwidth}
\caption{\label{Fig-CsF3} Bader charge of Cs in stable CsF$_3$ compounds as a function of pressure. Insets (a), $R\overline{3}m$ CsF$_3$ structure at 0 GPa; (b) $C$2/$c$ CsF$_3$ structure at 70 GPa.}
\end{figure}

Let us first look at CsF$_3$ compounds.
At ambient pressure, we find CsF$_3$ adopts a rhombohedral structure (space group $R\overline{3}m$), which is made of Cs$^+$ and linear symmetric [F$_3$]$^-$ species. 
The F-F distance in the [F$_3$] is 1.736 \AA, indicating as expected weaker bonding than in the F$_2$ molecule (F-F bond length 1.442 \AA).
Bader analysis also supports this conclusion:
there is a charge transfer of 0.950 $e$ from Cs to F$_3$ species, very close to the value in CsF (0.928 $e$),
but the charge distribution within F$_3$ is not even:
two end F atoms have the charge of -0.406 $e$, while the central atom only has -0.138 $e$.
Trihalide anions [X$_3$]$^-$ (X=Br, I) are well known. 
But, the [F$_3$]$^-$ species has only been experimentally found as CsF$_3$ complexes in argon matrix \cite{Riedel-IChem-2010, Ault-JACS-1976}. 
Here, we for the first time report its existence in a thermodynamically stable crystalline phase. 
According to DFT calculation, $R\overline{3}m$-CsF$_3$ is stable against decomposition to CsF and F$_2$ (the formation energy is about -0.189 eV/atom at T = 0 K, P = 1 atm).
At 27 GPa, CsF$_3$ undergoes a phase transition to a monoclinic phase $C$2/$c$. 
More interestingly, this structural transition coincides with a striking increase in Cs's Bader charge, as shown in Fig. \ref{Fig-CsF3}.
At 30 GPa, Cs has a charge of +1.7 $e$, far beyond the +1 valence state for alkali elements under ambient conditions, suggesting Cs$^{+}$ has been further oxidized.
Our previous study has shown that Xe under high pressure can be oxidized to +2, +4, +6 valence states. 
Cs$^{3+}$ is isoelectronic to Xe$^{2+}$.
This phase transition can be interpreted as a transition from Cs$^+$[F$_3$]$^-$ to [CsF$_2$]$^+$[F]$^-$.
This is also evidenced by the dramatic change of Cs(F)-F distance.
In the $C$2/$c$ phase, there are two types of F atoms (F1 and F2), each Cs is surrounded by 2 F1 and 4 F2.
At 50 GPa, the calculated Cs-F1 distance is 2.01 \AA, while the Cs-F2 distance is 2.58 \AA, consistent with a Jahn-Teller distortion related to the open-shell Cs$^{3+}$ configuration.
Compared with Cs-F bond length (2.69 \AA) in ionic CsF, we conclude that Cs-F2 interaction is very close to a typical ionic Cs-F bond, while each individual Cs-F1 bond has much stronger interaction (more covalent bonding).
Therefore, it can be viewed as [CsF$_2$]$^+$[F]$^-$ complex. 
A similar discussion can be also found in Miao's work \cite{Miao-NChem-2013}. 
But his previously proposed $C$2/$m$ structure is less stable than the $C$2/$c$ structure found here.

Similar to CsF$_3$, CsF$_5$ is also stable in the entire investigated pressure range between 0 - 100 GPa.
At 0 GPa, we found a monoclinic $P$2$_1$ phase is stable against decomposition to any other stable compositions (Cs, CsF, CsF$_3$, F).
$P$2$_1$-CsF$_5$ can be described as packing of Cs$^+$ and [F$_5$]$^-$ species.
[F$_5$]$^-$ ion has a V-shape and F-F bond lengths are 1.617, 1.953, 1.858, 1.617 \AA, and F-F-F bond angle at central F atom is 98.592$^\circ$.
Bader analysis shows that the entire F$_5$ group has charge -0.958 $e$. 
The hypothetical pentafluoride anion [F$_5$]$^{-}$ has also been proposed by Riedel \cite{Riedel-IChem-2010}, and we confirm it can exist at ambient pressure in a stable compound.
At 4 GPa, a new phase with [F$_5$]$^{-}$ groups and $C$2/$c$ symmetry becomes stable.  
Around 21 GPa, $C$2/$c$ phase transforms to another monoclinic $C$2/$m$ phase, which can be represented as [CsF$_2$] [F$_3$]. 
The [CsF$_2$] unit is very similar to the one in $C$2/$c$-CsF$_3$ (Bader charge is 0.820$e$), indicating that Cs achieves +3 state. 
At the same time, [F$_3$] is a typical polyfluoride anion with Bader charge -0.820$e$; thus the whole structure can be viewed as [CsF$_2$]$^+$[F$_3$]$^-$. 
At 47 GPa, consistent with Miao's results\cite{Miao-NChem-2013}, we found a structure based on the packing of CsF$_5$ molecules.
We again plot the variation of Cs's Bader charge in stable CsF$_5$ compounds with pressure. 
Indeed, analysis indicates a continuous two-step electronic transition of Cs, coinciding with the transition sequence (from $C$2/$c$ to $C$2/$m$ at 21 GPa, and from $C$2/$m$ to $Fddd$ at 47 GPa).
\begin{figure}
\epsfig{file=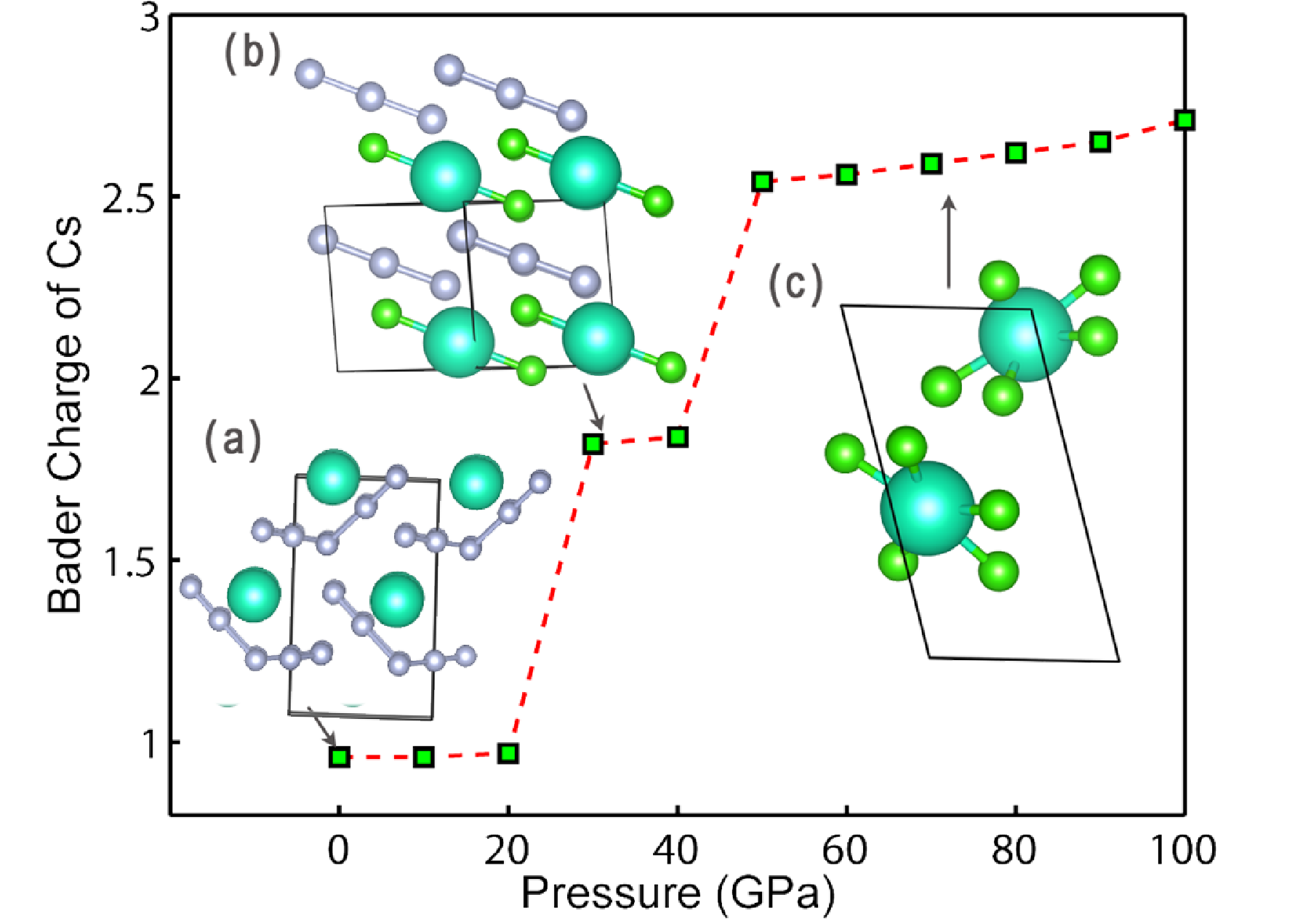, width=0.5\textwidth}
\caption{\label{Fig-CsF5} Bader charge of Cs in stable CsF$_5$ compounds as a function of pressure. Insets (a), $P$2$_1$-CsF$_5$ structure at 0 GPa; (b) $C$2/$m$-CsF$_5$ structure at 30 GPa; (c) $Fddd$-CsF$_5$ structure at 70 GPa}
\end{figure}

Our results show CsF$_3$ and CsF$_5$ are stable alongside the known compound CsF in the whole investigated pressure range (0-100 GPa). 
Unlike the recently discovered exotic sodium chlorides \cite{Zhang-Science-2013}, most of which are metallic, all of the predicted caesium fluorides are insulators.
There are two factors determining the stoichiometry of these insulating compounds: (1) Cs's valence state transition (+1 $\rightarrow$ +3 $\rightarrow$ +5); (2) formation of polyfluoride anions ([F$_3$]$^{-}$, [F$_5$]$^{-}$). 
Note that this is different from the previous study \cite{Miao-NChem-2013}, in which the latter factor was overlooked, along with a large number of stable phases.
Due to these two competing mechanisms, one can expect other stoichiometries can be stabilized as well. 
Indeed, we found CsF$_2$ and CsF$_4$ can be stable at intermediate pressure ranges.

\begin{figure}
\epsfig{file=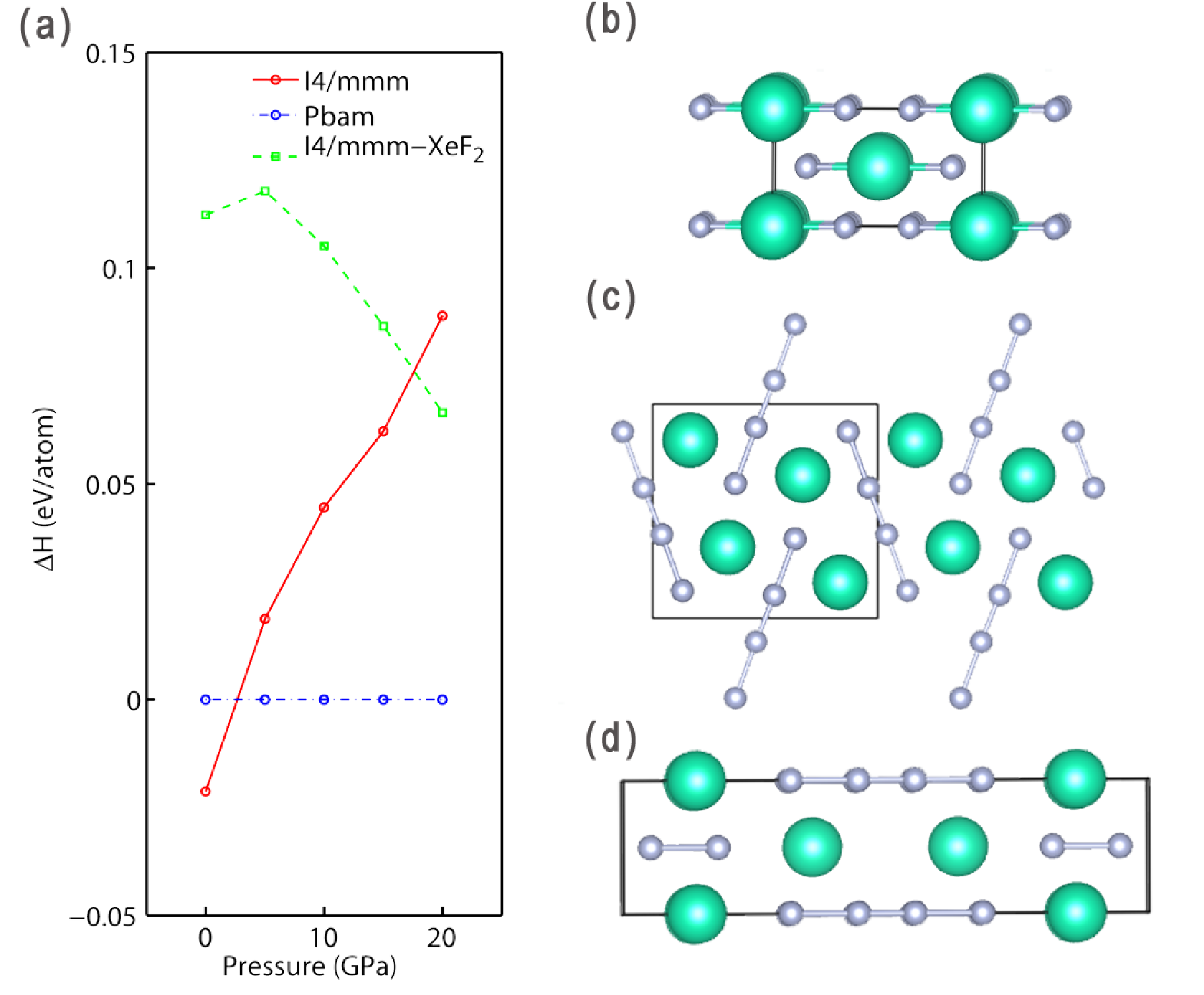, width=0.5\textwidth}
\caption{\label{Fig-CsF2} (a) Enthalpy of formation relative to $Pbam$-CsF$_2$ as a function of pressure; (b) unstable molecular $I$4/$mmm$-CsF$_2$ at 0 GPa; (c) stable $I$4/$mmm$-CsF$_2$ structure (stable between 0-2.8 GPa) ; (d) $Pbam$-CsF$_2$ structure (stable between 2.8-19 GPa).}
\end{figure}

Previously, a tetragonal ($I$4/$mmm$) XeF$_2$-like molecular structure was proposed to be stable at 5-20 GPa \cite{Miao-NChem-2013}. 
Our search found molecular CsF$_2$ crystal to be unstable against decomposition to CsF$_3$ and CsF at all pressures. 
A class of CsF$_2$ compounds, however, has been found to be stable at low pressures in our prediction. 
At 0 GPa, another $I$4/$mmm$ CsF$_2$ phase is found to be stable.
As shown in Fig. \ref{Fig-CsF2}c, it consists of [Cs]$^+$ and [F$_4$]$^{2-}$ ions.
The calculated Bader charges are 0.924 $e$ for [Cs], and -1.848 $e$ for [F$_4$], suggesting the formation of [Cs]$^+$ and [F$_4$]$^{2-}$.
Therefore, [Cs]$^{2+}$ is not favoured by energy, but CsF$_2$ can be stabilized due to the formation of the [F$_4$]$^{2-}$ anion.
[F$_4$]$^{2-}$ has not been observed by chemists so far, except that Riedel et al \cite{Riedel-IChem-2010} theoretically investigated the possibility of [F$_4$]$^-$. 
Yet our comprehensive structural search suggests that [F$_4$]$^{2-}$ based CsF$_2$ should be stable.
Tetraiodide anion [I$_4$]$^{2-}$ is known \cite{Polyiodide-2003}.
Our results suggest that fluorine follows that same trend under high pressure.
$I$4/$mmm$-CsF$_2$ would undergo a phase transition to an orthorhombic phase ($Pbam$) at 2.8 GPa, which also contains Cs$^+$ and [F$_4$]$^{2-}$ ions. 
Above 19 GPa, CsF$_2$ is no longer stable as there is a dramatic change in the valence of Cs from +1 to +3. 

\begin{figure*}
\epsfig{file=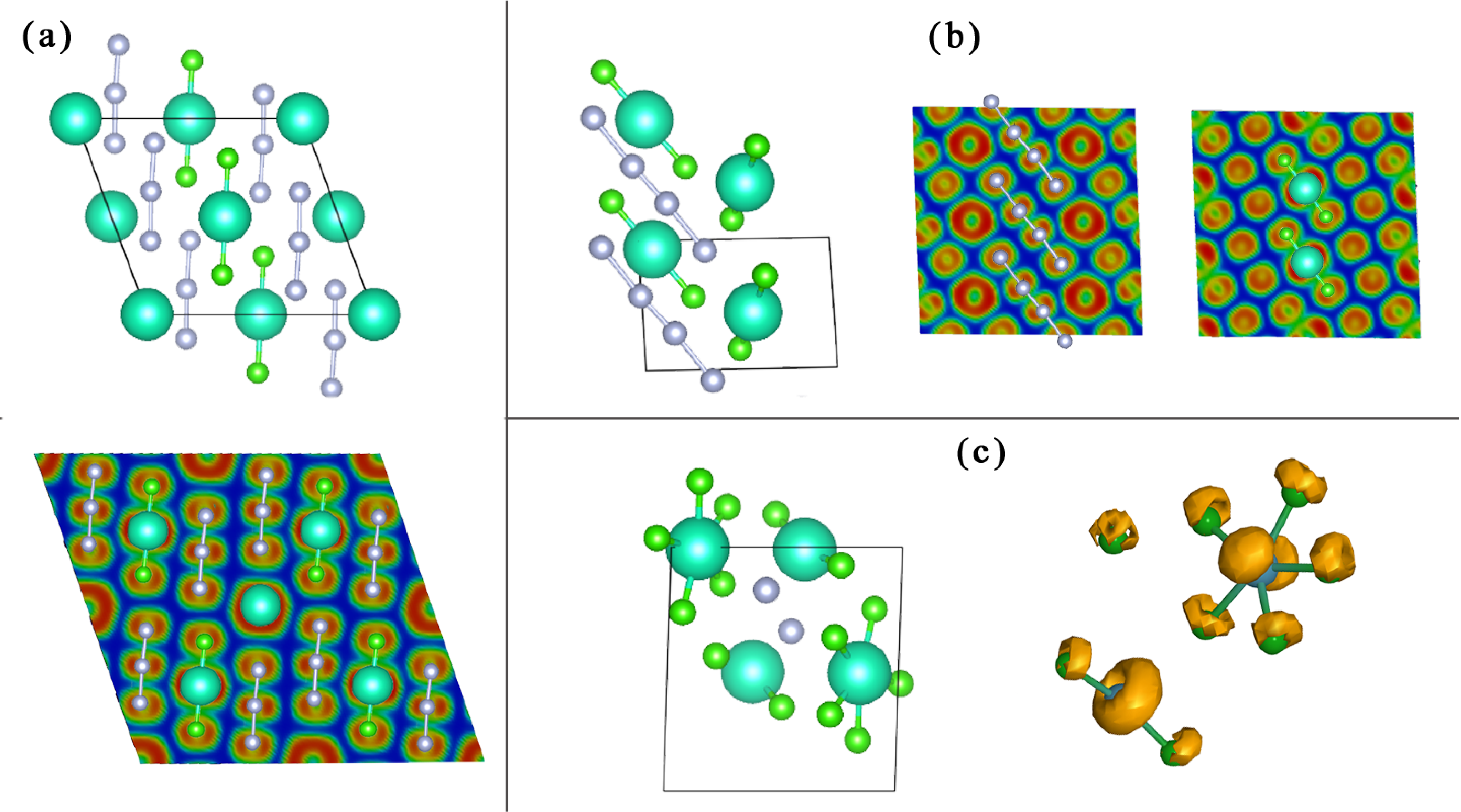, width=0.9\textwidth}
\caption{\label{CsF4} The stable crystal structures of CsF$_4$ and their corresponding (sliced or isosurfaced) ELF pictures at pressures of (a) 20 GPa, (b) 50 GPa, (c) 80 GPa.}
\end{figure*}

At around 20 GPa, CsF$_4$ becomes stable in a monoclinic form ($C$2/$m$) (Fig. \ref{CsF4}a). 
One can clearly see from the electron localization function (ELF) that half of Cs atoms have strong bonding with two neighbouring F atoms, and the other half of Cs atoms are simple Cs$^+$ cation.
The remaining F atoms form [F$_3$]$^-$ ions, as we already saw above in both CsF$_3$ and CsF$_5$.
Thus, it can be viewed as [CsF$_2$]$^+$[Cs]$^+$2[F$_3$]$^-$.
Bader analysis also supports this interpretation. 
Half of Cs atoms have Bader charge of 1.049$e$ , half of Cs have 1.773$e$. 
This suggests that Cs firstly achieves +3 valence state in CsF$_4$. 
At 31 GPa, all Cs atoms are oxidized to +3 state (Bader charge 1.829$e$), and the structure has space group $P$-1.
As shown in Fig. \ref{CsF4}b, the whole structure can be represented as 2[CsF$_2$]$^+$[F$_4$]$^{2-}$ ([F$_4$]$^{2-}$ anions appear again!). 
Miao investigated the possibility of CsF$_4$ molecule structurally similar to XeF$_4$, which is contradictory to chemical intuition (CsF$_4$ can be neither isostructural nor isoelectronic to XeF$_4$).
Indeed, our results suggest that Cs$^{4+}$ based molecule is energetically unfavored.
But we found that the most stable CsF$_4$ structure above 57 GPa shows valence state higher than +3. 
As shown in Fig. \ref{CsF4}c, it can be viewed as [CsF$_2$]$^{+}$[F]$^-$ $\cdot$ [CsF$_5$]$^0$.
Therefore, half of Cs atoms (in [CsF$_2$]) have +3 valence, while the other half of Cs (in [CsF$_5$]) have +5 valence state.
The resulting structure, crystallizing in $P$-1 symmetry, is stable up to 79 GPa between $C$2/$c$-CsF$_3$ and $Fddd$-CsF$_5$. 
We note that there also exists a stable phase of XeF$_3$ in the form of [XeF$_2$]$\cdot$ [XeF$_4$]\cite{XeF3-1963}.

Light halogens, fluorine (F) and chlorine (Cl), at normal conditions exist as highly reactive and toxic gases. 
For chemical industry and laboratory use, this presents great inconvenience. 
Their storage in the gaseous form (even as liquefied gases) is very inefficient, and compressed gas tanks may explode, presenting great dangers. 
At normal conditions, the volume of 22.4 litres (L) of pure fluorine gas weighs just 36 grams (g), illustrating the dismal inefficiency of storage in this form. 
To the best of our knowledge, no effective and safe fluorine storage materials are known. 
Both F and Cl have a huge range of industrial applications, which would benefit from such storage materials, especially if they can reach high storage capacity, stability and reversibility. 

In this work, we found that a series of CsF$_n$ ($n$=1, 2, 3, 5) compounds can be stable at zero temperature and ambient pressure. 
One mole of CsF$_5$ (227.9 g, occupying the volume of 0.07 L) contains 2 moles of F$_2$ gas (which in the free state would occupy the volume of 44.8 L - hence, storage in the form of CsF$_5$ is three orders of magnitude more efficient, and much safer, than in the form of pure F$_2$ gas). 
The reaction CsF$_5$ = CsF + 2F$_2$(gas), is thermodynamically unfavourable at zero temperature (the enthalpy of this reaction is 88.41 kJ/mol), but will be favourable on increasing temperatures, due to the higher entropy of the F$_2$ gas (202.8 J/(mol$\cdot$K) at standard conditions) \cite{chemistry-2006}. 
The calculated thermodynamic properties of these defluorination reactions are given in Table \ref{reactions}. 
It can be seen that such compounds as CsF$_3$ can be thermally decomposed, and then again be synthesized at lower temperatures at nearly room temperature window. 
CsF$_5$, having the highest F content, can be used for fluorine storage at low temperature conditions.
Such reversibility is a great advantage of the proposed fluorine storage materials.

\begin{table} [!htb]
\begin{center}
\caption{\label{reactions}Investigated reactions of the CsF-F system at ambient pressure conditions. wt\% gives the weight content of released F$_2$ gas. $\Delta$H$^{\textrm 0K}$ and $\Delta$H$^{\textrm 300K}$ are the calculated enthalpies at $T$=0 K and 300 K, including the vibrational energies in (kJ/mol). $\Delta$S$^{\textrm 300K}$ is the corresponding formation entropy in J/(K$\cdot$mol). $T_c$ is the predicted decomposition temperature at standard atmosphere (1 bar). Note that F$_2$ is treated as the crystalline solid at 0 K. }
\begin{tabular}{lccccc}
\hline
\hline

Reactions  &  wt \% &$\Delta$H$^{\textrm 0K}$ & $\Delta$H$^{\textrm 300K}$ & $\Delta$S$^{\textrm 300K}$ & $T_c$($^\circ$C) \\
\hline
CsF$_2$  $\rightarrow$  CsF + $\frac{1}{2}$F$_2$(g)  &  11.1  & 44.30 & 37.59 &  78.25 & 218 \\
CsF$_3$  $\rightarrow$  CsF + F$_2$(g)               &  20.0  & 72.24 & 63.41 & 152.29 & 150 \\
CsF$_5$  $\rightarrow$  CsF + 2F$_2$(g)              &  33.3  & 88.41 & 76.73 & 284.96 & -15 \\
\hline
\hline

\end{tabular}
\end{center}
\end{table}

\begin{figure}
\epsfig{file=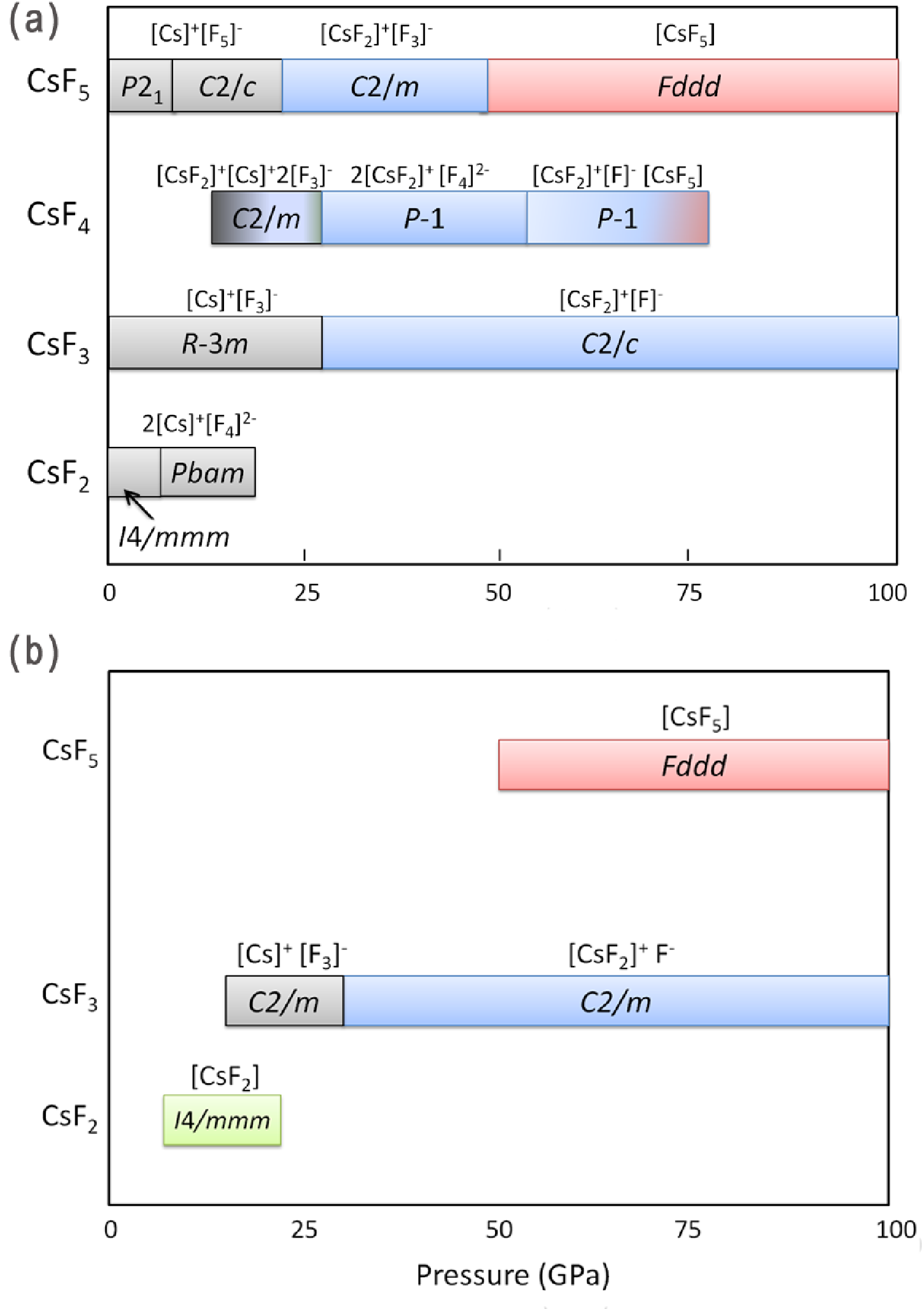, width=0.48\textwidth}
\caption{\label{Phase}Comparison of CsF$_n$ ($n$=2,3,4,5) stability phase diagram with respect to pressure. (a) revised results from this study; (b) results from previous literature \cite{Miao-NChem-2013}. Note that that each colour represent distinct Cs's valence state in the given compounds (grey: Cs$^+$; green: Cs$^{2+}$; blue: Cs$^{3+}$; red: Cs$^{5+}$), while the gradient colour indicate a mixed valence states in between. The $I$4/$mmm$-CsF$_2$ structure in (a) and (b) are very different. The only common phase to (a) and (b) is $Fddd$-CsF$_5$.}
\end{figure}

We have presented a comprehensive study of possible stable compounds in the CsF-F binary system under pressure. 
CsF$_n$ phases show extremely rich chemistry. 
At ambient pressure, a series of compounds CsF$_2$, CsF$_3$, CsF$_5$ are thermodynamically stable because of the formation of polyfluoride anions of [F$_3$]$^{-}$, [F$_4$]$^{2-}$, [F$_5$]$^{-}$. 
Our results confirm the previously proposed polyfluoride anions (F$_3$$^-$, F$_5$$^-$), and a new ion (F$_4$$^{2-}$). 
Under high pressure, 5$p$ electrons in Cs atoms can become chemically active, making Cs$^{3+}$ and Cs$^{5+}$ energetically favourable. 
Although our prediction found Cs$^{2+}$ and Cs$^{4+}$ states are far from being stable, the stochiometric compounds CsF$_2$ and CsF$_4$ can become stable at 0-20 and 15-80 GPa, respectively, but these contain Cs$^+$, Cs$^{3+}$, Cs$^{5+}$, [F$_4$]$^{2-}$, [F$_3$]$^-$.
As shown in Fig. \ref{Phase}, crystal structures of caesium polyfluorides can be summarized as the packing between Cs-containing cations (Cs$^{+}$, [CsF$_2$]$^{+}$), polyfluoride anions ([F$_3$]$^{-}$, [F$_4$]$^{2-}$, [F$_5$]$^{-}$), and neutral molecular species (CsF$_5$).
We hope this report will stimulate further experimental studies and serve as a guide for the design of fluorine storage materials.

\section{Methods}
Searches for the stable compounds and structures were performed using an evolutionary algorithm, as implemented in the USPEX code \cite{Oganov-JCP-2006, Oganov-ACR-2011, Lyakhov-CPC-2013, Zhu-Acta-2012, Zhu-PRB-2013}. The most significant feature of USPEX we used in this work is the capability of optimizing the composition and crystal structures simultaneously - as opposed to the more usual structure predictions at fixed chemical composition \cite{Zhu-MgO-PCCP, Hu-PRL-2013, Zhang-Science-2013}. 
The compositional search space is described via building blocks (for example, search for all compositions in a form of [$x$CsF + $y$F]). During the initialization, USPEX samples the whole range of compositions of interest randomly and sparsely. Chemistry-preserving constraints in the variation operators are lifted and replaced by the block correction scheme which ensures that a child structure is within the desired area of compositional space, and a special ``chemical transmutation'' operator is introduced. Stable compositions are determined using the convex hull construction: a compound is thermodynamically stable if the enthalpy of its decomposition into any other compounds is positive. Structure prediction was done in conjunction with ab initio structure relaxations based on density functional theory (DFT) within the Perdew-Burke-Ernzerhof (PBE) generalized gradient approximation (GGA) \cite{PBE} as implemented in the VASP code \cite{VASP}. For structural relaxation, we used the all-electron projector-augmented wave (PAW) method and the plane wave basis set with the 600 eV kinetic energy cutoff; the Brillouin zone was sampled by Monkhorst-Pack meshes with the resolution 2$\pi$ $\times$ 0.06 \AA$^{-1}$. For post-processing, the selected low-enthalpy structures were treated by using hard PAW potential of F (F\_h), using a energy cut off of 1000 eV.
Such calculations provide an excellent description of the known structures (CsF and F$_2$) and their energetics. To ensure that the obtained structures are dynamically stable, we calculated phonon frequencies throughout the Brillouin zone using the finite-displacement approach as implemented in the Phonopy code \cite{phononpy-2008}.
The vibrational entropies and enthalpies are obtained by directly summing over the calculated phonon frequencies, in order to calculate the free energy (see online supporting information, similar methods have been widely used for simulation of dehydration reactions for hydrogen storage materials \cite{Wolverton-PRB-2004, Ozolins-JACS-2009}).
Charge transfer was investigated on the basis of the electron density using Bader's analysis \cite{Bader-1990} as implemented in a grid-based algorithm without lattice bias \cite{Tang-JPCM-2009}.
Electron localization functions (ELF) \cite{Becke-JCP-1990} are also calculated in order to analyze chemical bonding for the selected compounds.

\begin{acknowledgements}
This work is funded by DARPA (Grants No. W31P4Q1210008 and No. W31P4Q1310005), the NSF (No. EAR-1114313 and No. DMR-1231586), the Basic Research Foundation of NWPU (Grant No. JCY20130114), the Natural Science Foundation of China (Grants No. 51372203 and No. 51332004), the Foreign Talents Introduction and Academic Exchange Program (Grant No. B08040). Calculations were performed on the supercomputer of Center for Functional Nanomaterials, Brookhaven National Laboratory, which is supported by the U.S. Department of Energy, Office of Basic Energy Sciences, under contract No. DE-AC02 98CH10086. The authors also acknowledge the High Performance Computing Center of NWPU for
the allocation of computing time on their machines.
ARO thanks V. Mukhanov for the idea of fluorine storage applications.
\end{acknowledgements}

%

\end{document}